\documentclass[11pt,a4paper]{article}

\usepackage{latexsym}
\usepackage[dvips]{color}
\usepackage{graphicx}

\newcommand{\hb}{{\textit{hb}}}

\begin{document}

\title{The competition of hydrogen-like and isotropic interactions on polymer collapse}

\author{J.\ Krawczyk$^1$, A. L. Owczarek$^1$ and T. Prellberg$^2$\thanks{{\tt {\rm email:}
j.krawczyk@ms.unimelb.edu.au,aleks@ms.unimelb.edu.au,t.prellberg@qmul.ac.uk}} \\
         $^1$Department of Mathematics and Statistics,\\
         The University of Melbourne,\\
         Parkville, Victoria 3052, Australia.\\
$^2$School of Mathematical Sciences\\
Queen Mary, University of London\\
Mile End Road, London E1 4NS, UK
}

\maketitle

\begin{abstract}
  We investigate a lattice model of polymers where the
  nearest-neighbour monomer-monomer interaction strengths differ
  according to whether the local configurations have so-called
  ``hydrogen-like'' formations or not.  If the interaction strengths
  are all the same then the classical $\theta$-point collapse
  transition occurs on lowering the temperature, and the polymer
  enters the isotropic liquid-drop phase known as the collapsed
  globule. On the other hand, strongly favouring the hydrogen-like 
  interactions give rise
  to an anisotropic folded (solid-like) phase on lowering the
  temperature. We use Monte Carlo simulations up to a length of 256 to map
  out the phase diagram in the plane of parameters and determine the
  order of the associated phase transitions. We discuss the
  connections to semi-flexible polymers and other polymer models.
  Importantly, we demonstrate that for a range of energy parameters
  two phase transitions occur on lowering the temperature, the second
  being a transition from the globule state to the crystal state. We
  argue from our data that this globule-to-crystal transition is continuous 
  in two dimensions in accord with field-theory arguments concerning
  Hamiltonian walks, but is first order in three dimensions.
\end{abstract}

\newpage

\section{Introduction}
The self-avoiding walk (SAW) on a lattice
\cite{Flory:book,deGennes:book} is a key model in statistical
mechanics for the study of the static properties of polymers.
Incorporating interactions in this model make it possible to represent
many features of real polymers.  Regardless of the constraints of the
lattice relative to the real world, the model mimics very well many
properties of physical systems \cite{Vanderzande:book}.  The
self-avoiding walk on a lattice is a random walk which is not allowed
to visit a lattice site more then once. Each visited lattice site is
considered as a monomer of the polymer chain. A common way
\cite{deGennes:book} to model intra-polymer interactions in such a
walk is to assign an energy to each non-consecutive pair of monomers
lying on the neighbouring lattice sites. This is the canonical ISAW
model which is the standard model of polymer collapse
using self-avoiding walks. With this modification one studies a
polymer in a solvent, where the energy between monomers can be
attractive or repulsive and depends on temperature.  If the energy is
repulsive the polymer behaves as a swollen chain (the so-called
excluded-volume state) regardless of temperature and one says that it
is in a good solvent. When the energy is attractive, and the
temperature is low enough, the chain becomes a rather more compact
globule \cite{deGennes:book,desCloiseaux:book}, reminiscent of a
liquid droplet: this is also known as the poor solvent situation. The
transition point between those two phases is called the $\theta$-point;
it is a well studied continuous phase transition (see
\cite{prellberg:1994-01} and references therein).

The modelling changes as soon as we want to describe any biological
system (e.g.  proteins), in which the hydrogen bonding plays an
important role \cite{pauling:1951-01}. One of the main features of the
bonding is that the interacting residua lie on a partially straight
segments of the chain.  Hydrogen-like bonding was first modelled on
the cubic and square lattices using Hamiltonian paths by Bascle \textit{et al.\ }
\cite{bascle:1993-01}.  A monomer acquires a hydrogen-like bond
with its (non-consecutive) nearest neighbour if both of them lie on
straight sections of the chain (see Figure~\ref{fig_interactions}).  The
interacting self-avoiding walk modified to have only such interactions
will be referred to as the hydrogen-like bonding model, or rather
\textit{hb}-model.  The \textit{hb}-model was studied in mean-field
approximation \cite{bascle:1993-01} and a first-order transition from
a high-temperature excluded-volume (swollen) phase to a quasi-frozen
solid-like phase was found in both two and three dimension.  Hence
this would indicate that it is a different type of transition from the
$\theta$-point. Note also that the low temperature \textit{hb}-phase
is anisotropic whereas the collapsed globule of the standard
$\theta$-point model is isotropic.  The 
\textit{hb}-model on the square lattice was studied directly by Foster and Seno
by means of the transfer matrix method \cite{foster:2001-01} 
and by Krawczyk \textit{et al.\ }\cite{my:2007-01} on both the
square and cubic lattice using a Monte Carlo method.  In both studies a
first-order transition was found between an excluded-volume
(swollen-coil) state and an anisotropic ordered compact phase in two
and in three dimensions, again in opposition to the $\theta$-point
\cite{deGennes:book}.

It is appropriate to compare this difference between the behaviour of
the \textit{hb}-model and $\theta$-point models with the difference
between the behaviour of interacting semi-stiff polymers and the
fully-flexible $\theta$-point polymers. This is because hydrogen
bonding induces an effective stiffness in the polymer between those
monomers that are taking part in the interactions. As the temperature
is lowered the proportion of the monomers experiencing this stiffness
increases so while not all the segments of the polymer feel this
stiffness at high temperatures, the proportion of monomers involved
with nearest-neighbour \hb-interactions increases towards unity as the
temperature is lowered.  In three dimensions, Bastolla and Grassberger
\cite{bastolla:1997-01} discussed so-called semi-stiff self-avoiding
walks, which interact via all nearest-neighbours, as in the
$\theta$-point model, and include a bending energy. They showed that
when there is a strong energetic preference for straight segments,
this model undergoes a single first-order transition from the
excluded-volume high-temperature state to a state similar to the
low-temperature solid-like state of the \textit{hb}-model.
Intriguingly, if there is only a weak preference for straight
segments, the polymer undergoes two phase transitions: on lowering the
temperature the polymers undergoes the $\theta$-point transition to
the liquid globule followed at a lower temperature by a first-order
transition to the frozen phase. We should point out though that in two
dimensions the transition between the globule and the frozen state has
only been studied in Hamiltonian walks, and there it seems to be
continuous one \cite{jacobsen:2004-01}.

To complicate matters further, there is at least one other model using
a different definition of interactions which could be regarded as
hydrogen-like bonding.  This model \cite{buzano:2002-01} defines
interactions between parallel segments, that is, bonds of the lattice
occupied by the walk and so connecting monomers, see
Figure~\ref{fig_interactions}. We will call this model the interacting
bond model. Studying this model by means of Bethe approximation,
Buzano and Pretti \cite{buzano:2002-01} found, in both two and three
dimensions, two phase transitions: while decreasing the temperature
the $\theta$-collapse to isotropic globule phase is followed by a
first order transition to a solid-like phase. Hence, this is similar
to the semi-stiff model for weak stiffness.  The interacting-bond
model in two dimensions has recently been studied by Foster
\cite{foster:2007-01} and also displays two transitions. In
\cite{buzano:2003-01} Buzano and Pretti added isotropic
nearest-neighbour monomer-monomer interactions to the interacting-bond
model, and investigated the phase diagram in three dimensions, again
in the Bethe approximation. They showed that the phase diagram is
similar to the interacting-bond model.  However, if the interaction
between monomers are repulsive, there is only one first-order phase
transition from the swollen coil to the solid-like phase. This is
again reminiscent of the semi-stiff model for strong stiffness.

It is therefore of some interest to study an enhanced
\textit{hb}-model where non-hydrogen bond nearest neighbours are also
considered. Hence, in this paper we investigate a model of
self-interacting self-avoiding walks with two types of
nearest-neighbour interaction: the \textit{hb}-interactions and
nearest-neighbour interactions that are \emph{not} hydrogen bonds,
which we denote as \textit{nh}-interactions.  The competition between
these two types of interaction (\textit{hb} versus \textit{nh}) leads
to a three-phase phase diagram, with excluded volume, globule and
frozen phases. Of special interest is the comparison to the semi-stiff
model.  One key question is whether there can exist two phase
transitions on lowering the temperature. The order of the transitions
in two and three dimensions are also of interest. We use a Monte Carlo
technique, known as FlatPERM \cite{my:2004-01}, to study self-avoiding
walks on the simple cubic and square lattices with interactions as
described.

The paper is organised as follows. In Section 2 we explain more
carefully details of the model.  In Section 3 the phase diagram in
both dimensions is discussed. A discussion of the anisotropy of the
model is also given.  We conclude with a summary and discuss the
similarity of this model to the semi-stiff model.

\section{Model and simulations}

\begin{figure}[ht!]
\center{
\includegraphics[scale=0.5,angle=0]{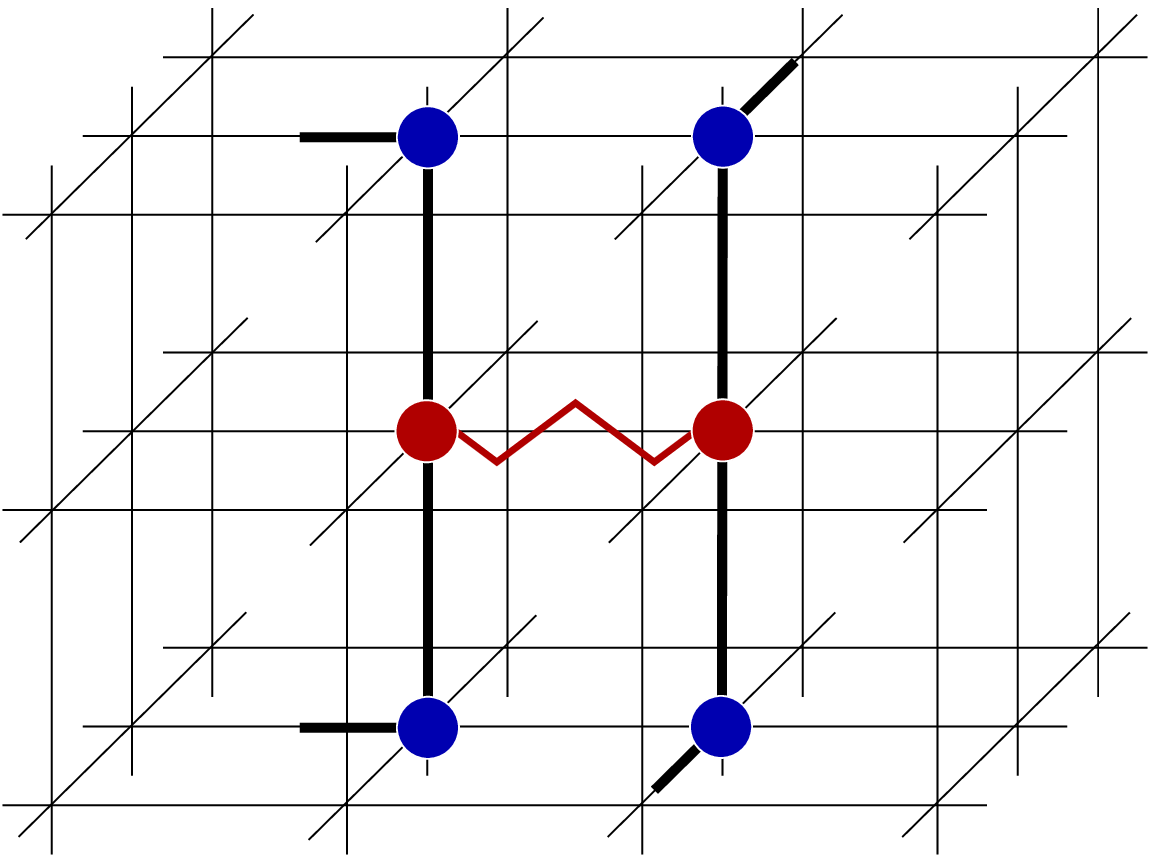}
\hspace{0.5cm}
\includegraphics[scale=0.5,angle=0]{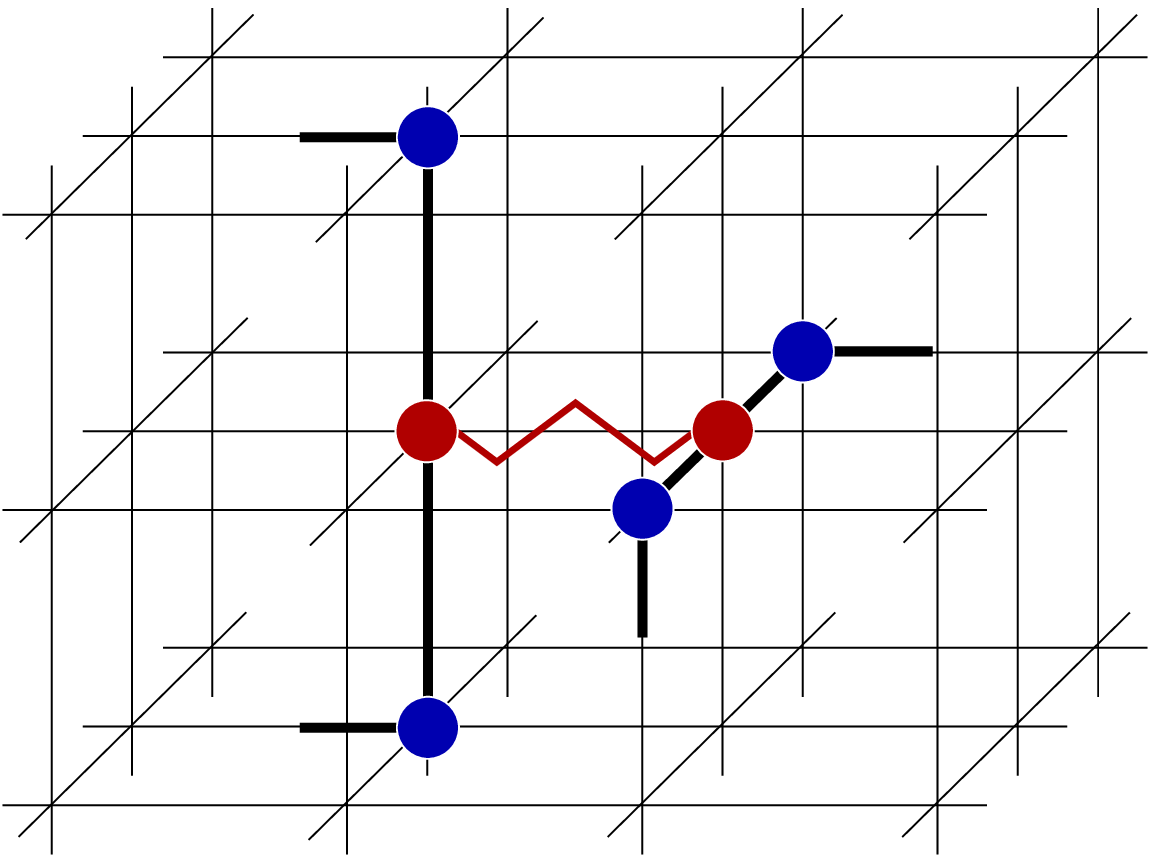}
}
\caption{\label{hb_interactions} The types of nearest-neighbour
  interactions between two straight segments of the polymer involved
  in the \textit{hb}-interactions: parallel segments (left) and
  orthogonal segments (right). In the model studied in this paper
  these two types are weighted equally.  In two dimensions, only
  parallel interactions are possible.}
\end{figure}

The polymer is modelled on a square and simple cubic lattice as a
self-avoiding walk with interactions between different types of
nearest-neighbours monomers: that is monomers that are not consecutive
in the walk but nearest neighbours on the lattice.  The strength of
the interaction depends on the relative position of the monomers
involved in the interaction to those next to them on the walk.  A
segment is defined as a site along with the two adjoining bonds
visited by the walk, and we say that a segment is straight if these
two bonds are aligned. The `hydrogen bonds' are nearest neighbour
interactions that are between monomers where both are part of
\emph{straight} segments of the polymer, see
Figure~\ref{hb_interactions}.

\begin{figure}[ht!]
  \centering
  \includegraphics[scale=0.7]{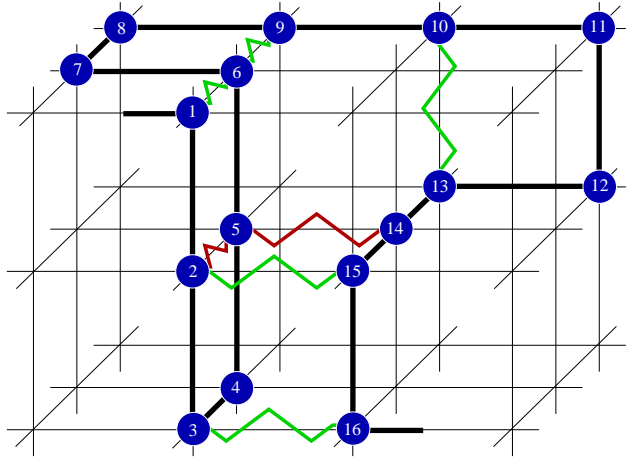}
  \caption{Interactions in $3d$ interacting self-avoiding walk. In our
model we distinguish two different kind of interactions, which are
denoted by two different colours. The red colour (dark shading)
denotes hydrogen-like interactions (\textit{hb}) (monomers $2$-$5$,
$5$-$14$) whereas the green colour (light shading) denotes all other
interactions between two neighbouring monomers (\textit{nh}) (monomers
$1$-$6$ and $10$-$13$ for example).}
  \label{fig_interactions}
\end{figure}

Our model weights both the parallel and orthogonal \textit{hb}
interactions equally. Our model also includes all other possible
nearest neighbour interactions and assigns them a different Boltzmann
weight. The two types of distinguished interactions are shown in
Figure~\ref{fig_interactions}.  As just described interactions between
monomers sitting on straight lines, as in monomer $\#5$ of the triple
of monomers $\#2$, $\#5$ and $\#14$, form the hydrogen-bond
interactions with like monomers, as in the interacting pairs of
monomers $2$-$5$ and $5$-$14$. The energy of those interactions is
denoted as $-\varepsilon_{hb}$.  The other kind of interaction consist
of any nearest-neighbour interaction between non-consecutive monomers
that are not hydrogen bonds, as in the pairs $1$-$6$ and $10$-$13$ for
example. The energy of those interactions is denoted as
$-\varepsilon_{nh}$.

Note that the number of all nearest-neighbours interactions, $m$, is
equal to the sum of the number of the two types of interaction
considered in our model, that is $m=m_{nh}+m_{hb}$.  The energy of
configuration $\varphi_n$ of an $n$-step walk is calculated as
\begin{equation}
E(\varphi_n)=-m_{hb}(\varphi_n)\cdot\varepsilon_{hb}-m_{nh}(\varphi_n)\cdot\varepsilon_{nh}\;,
\end{equation}
where $m_{hb}$ and $m_{nh}$ are the number of hydrogen-like bond
interactions and non hydrogen-like nearest-neighbour interactions,
respectively.  The inverse temperature is denoted as $\beta=1/k_{B}T$,
where $k_B$ is the Boltzmann constant and $T$ the absolute
temperature.  We define for convenience
$\beta_{hb}=\beta\varepsilon_{hb}$ and
$\beta_{nh}=\beta\varepsilon_{nh}$.  The partition function is then
given by
\begin{equation}
Z_n(\beta_{hb},\beta_{nh})=
\sum_{m_{hb},m_{nh}} C_{n,m_{hb},m_{nh}}\; e^{\beta_{hb} m_{hb}+\beta_{nh} m_{nh}}
\end{equation}
with $C_{n,m_{hb},m_{nh}}$ the density of states. Canonical averages
are calculated with respect to this density of states.

Since we will consider simulational results along lines in parameter
space at a constant ratio of $\varepsilon_{nh}/\varepsilon_{hb}=
\beta_{nh}/\beta_{hb}$ we define $\varepsilon_{nh}=\gamma$, and
$\varepsilon_{hb}=1-\gamma$, so that the energy is then given by
\begin{equation}
E=-m_{hb}\cdot(1-\gamma)-m_{nh}\cdot\gamma.
\end{equation}
In our study we will analyse the (reduced) specific heat to
investigate the phase diagram: that is,
\begin{equation}
C(T)=\frac{1}{T}\frac{\langle E^2\rangle-\langle E\rangle^2}{n}.
\end{equation}

We will also consider fixing one of the parameters either $\beta_{hb}$
or $\beta_{nh}$ and varying the other. To analyse the possible phase
transitions we then use the fluctuations in number of monomers of
appropriate type. In the case of $\beta_{hb}$ being constant we
consider
\begin{equation}
\sigma^2(m_{nh})=\langle m^2_{nh}\rangle-\langle m_{nh}\rangle^2.
\end{equation}
When $\beta_{nh}$  is constant we consider 
\begin{equation}
\sigma^2(m_{hb})=\langle m^2_{hb}\rangle-\langle m_{hb}\rangle^2.
\end{equation}

Simulations have been performed with FlatPERM algorithm
\cite{my:2004-01}. We have simulated the models using a two parameter
implementation (utilising $m_{hb}$ and $m_{nh}$) for length $n=128$
where the simulation directly estimates this density of states
$C_{n,m_{hb},m_{nh}}$.  We have also performed one parameter ($m_{hb}$
or $m_{nh}$) simulations for systems of size $256$ where the simulation
estimates partial summations of this density of states over one of the
variables.

\section{Results and discussion}

\subsection{Pure hydrogen bonding and the canonical ISAW models}

When $\gamma=1/2$ the model becomes the canonical interacting self
avoiding walk (ISAW), which displays the $\theta$-transition from coil
to globule state. The $\theta$-transition is a second-order phase
transition in both two and three dimensions.  In two dimensions the
established crossover exponent is $\phi=3/7$ \cite{duplantier:1987},
which implies a negative specific heat exponent $\alpha=2-1/\phi=-1/3$
(the specific heat does not diverge on approaching the transition)
\cite{brak:1993-01}.  In three dimensions, which is the upper critical
dimension for the $\theta$-transition, the specific heat is expected
to diverge logarithmically \cite{duplantier:1986}.  The collapsed
state is an isotropic dense liquid-like droplet with a well-defined
surface tension \cite{owczarek:1993,baiesi:2006}.

On the other hand, for $\gamma=0$ the model becomes the
\textit{hb}-model studied by Foster and Seno \cite{foster:2001-01} on
the square lattice and Krawczyk \emph{et al.\ }on the square and
simple cubic lattices \cite{my:2007-01}. In both two and three
dimensions, there is a single first-order transition to a folded
crystalline state, which is anisotropic.

Using these results as a starting point, one therefore expects there
to be at least these three phases (swollen, globule, crystal) in the
full two-dimensional parameter space explored here.  For $\gamma>1/2$,
the \hb-interactions are suppressed relative to the non-\hb\
interactions, and the simplest hypothesis would be that the
$\theta$-transition is not affected. To test this hypothesis, we have
considered the line $\gamma=1$ below.  For $\gamma$ close to zero, the
\hb-interactions dominate, and one may expect that there exists some
range of values for $\gamma$ for which the first-order transition of
the pure \hb-model persists.  To test this, we need to consider a
small value of $\gamma$.  We of course then need to consider other
values of $\gamma$ to see if the two transitions can occur for fixed
$\gamma$ and whether there exist any other phases.

\subsection{Features of the phase diagram}

\begin{figure}[ht!]
  \centering
  \includegraphics[width=0.7\textwidth]{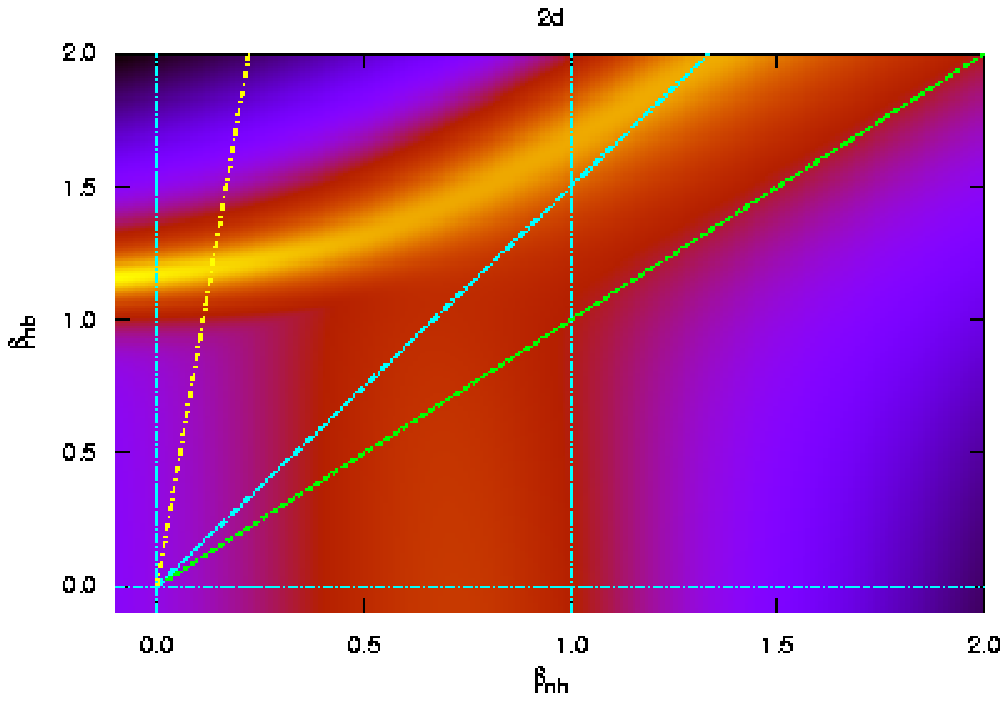}
   \includegraphics[width=0.7\textwidth]{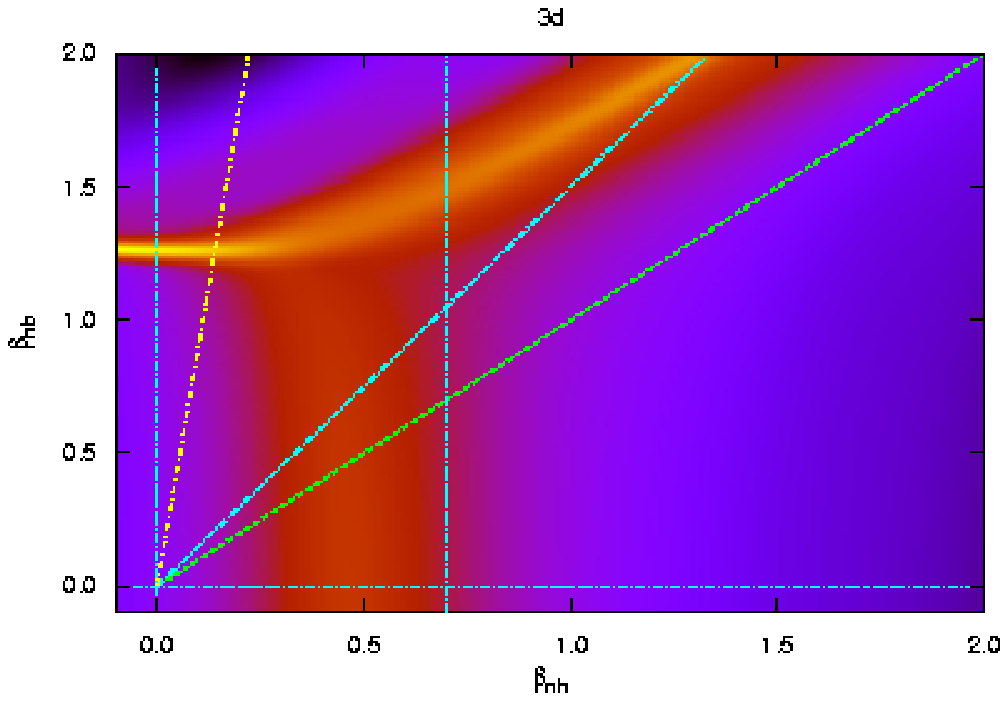}
   \caption{Density plots of the logarithm of the largest eigenvalue
of the matrix of second derivatives of the free energy with respect to
$\beta_{nh}$ and $\beta_{hb}$ (the lighter the shade the larger the
value).  Both plots are for $n=128$, for two- and three-dimensional
systems, top and bottom, respectively.  The lines shown indicate
cross-sections for which we have performed additional simulations or
analysis.  In both pictures we show lines with slope
$(1-\gamma)/\gamma$ for $\gamma=0.0, 0.1, 0.4, 0.5$ and
$1.0$. Displayed are also vertical lines at $\beta_{nh}=1.0$ and $0.7$
in two and three dimensions, respectively.
\label{fig_lambda}}
\end{figure}

To gain an understanding of which values of $\gamma$ we may need to
consider more closely, we first examine the fluctuations in the
numbers of interaction across a wide range of
$(\beta_{hb},\beta_{nh})$.  As in previous work
\cite{krawczyk2005a,krawczyk2005b,my:2007-01}, we found the use of the
largest eigenvalue of the matrix of second derivatives of the free
energy with respect to the parameters $\beta_{nh}$ and $\beta_{hb}$
most advantageous to show the fluctuations in a unified manner.
Figure~\ref{fig_lambda} displays density plots of the size of
fluctuations for $-0.1\leq\beta_{nh},\beta_{hb}\leq 2.0$ in two and
three dimensions, respectively.  The lighter the shade the larger the
fluctuations.

It suggests the presence of three thermodynamic phases separated by
three phase transition lines meeting at a single point. From our
discussion above, we can therefore identify these three phases as
swollen, globule, and crystal.  Therefore, for small values of
$\beta_{nh}$ and $\beta_{hb}$, we expect the model to be in the
excluded volume universality class of swollen polymers, since at
$\beta_{nh}=\beta_{hb}=0$ the model reduces to simple self-avoiding
walks.  For fixed $\beta_{hb}$ and large $\beta_{nh}$, we deduce that
the polymer is in a globular state, while for fixed $\beta_{nh}$ and
large $\beta_{hb}$, we deduce that the polymer is in the anisotropic
crystalline state.

A fixed value of $\gamma$ corresponds to a straight line of slope
$(1-\gamma)/\gamma$ on the plots in Figure~\ref{fig_lambda}.  The line
with $\gamma=0.5$ is shown on both plots and it is clear that for any
value of $\gamma\geq0.5$ (slope less than $1$) the system will undergo
only one phase transition on lowering the temperature. This transition
should be in the universality class of the $\theta$-transition. Below
we consider the line $\gamma=1$ to verify this.  The line with
$\gamma=0.1$ is shown on both plots and it is clear that there is
indeed a range of values of $\gamma$ around $0$ (slope sufficiently
large) for which the system will undergo a single first-order
\hb-model like transition on lowering the temperature.  The figures
also suggest that there exists a critical value of $\gamma$, say,
$\gamma_c$, where this scenario ends.

Therefore one may deduce that there exists some range of $\gamma$
between $\gamma_c$ and $0.5$, not necessarily the whole range, for
which the system undergoes two transitions on lowering the
temperature. In this region, the polymer starts in the swollen state
at high temperatures, undergoes a $\theta$-transition to a globular
state on lowering the temperature, and on lowering the temperature
further, undergoes a further (novel) transition to the crystalline
state.  We verify this by considering the line with $\gamma=0.4$.

\begin{figure}[ht!]
  \centering
  \includegraphics[width=0.7\textwidth]{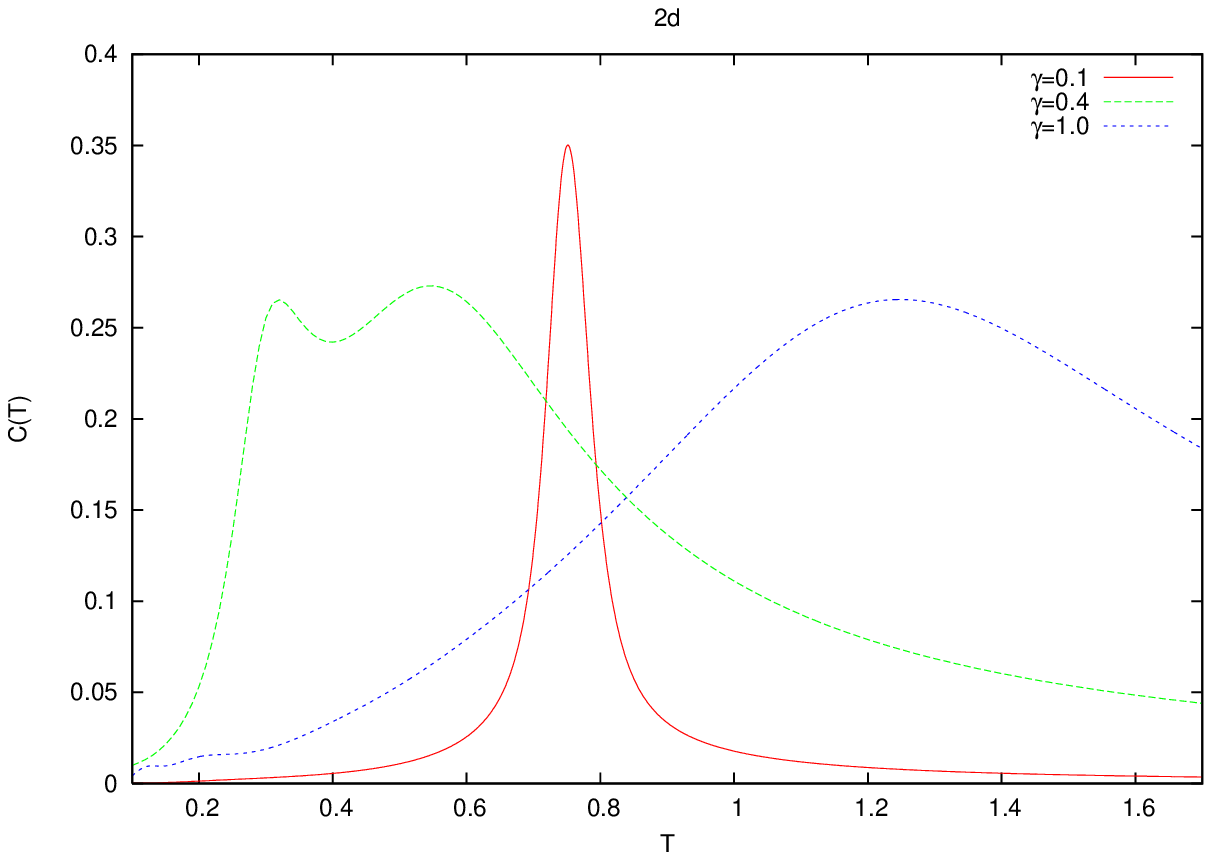}
  \includegraphics[width=0.7\textwidth]{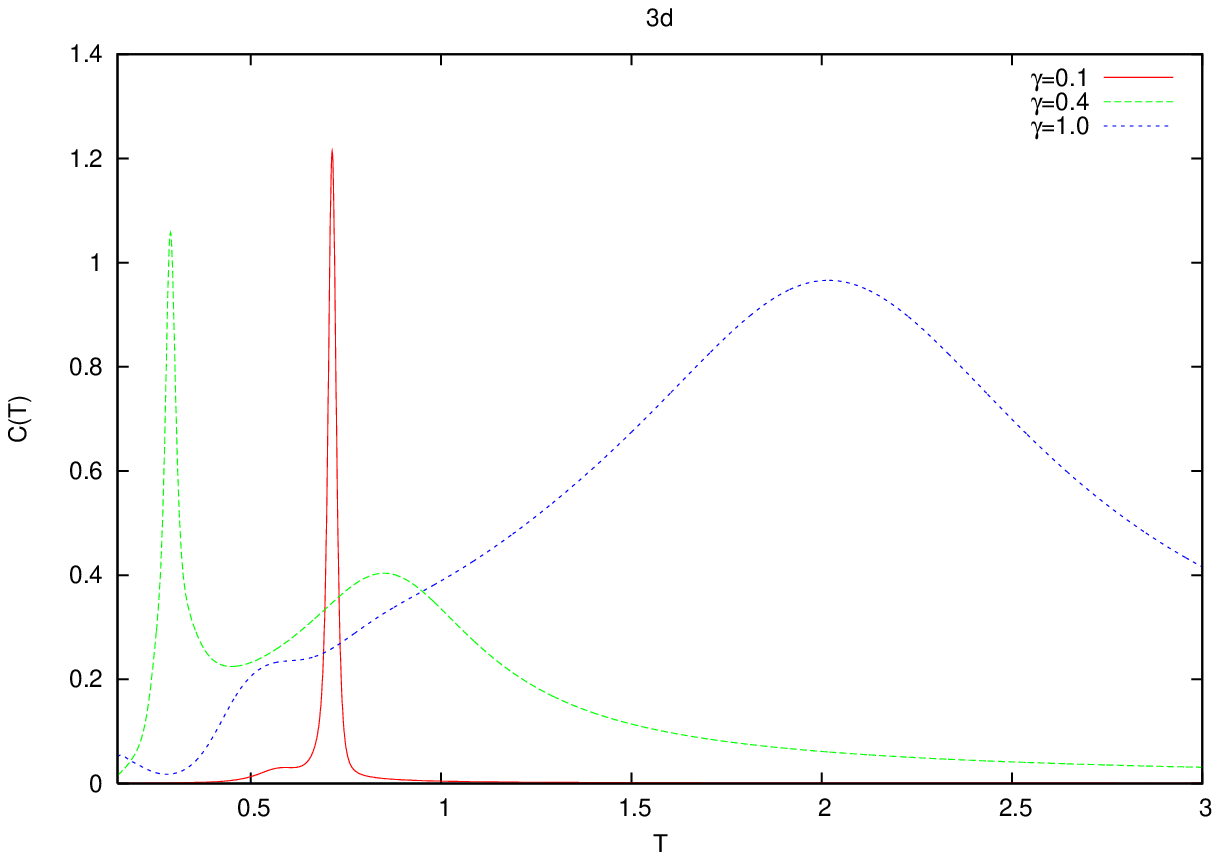}
  \caption{The specific heat for different values of $\gamma$ in two
    and three dimensions for $n=128$.  For $\gamma=0.1$ and $1.0$ we
    see only one maximum in the specific heat, so we expect only one
    phase transition.  For $\gamma=0.4$ two maxima occur: upon
    changing the temperature there are two phase transitions.  (The
    specific heat for $\gamma=0.1$ is divided by a factor $15$ in
    three dimensions and $10$ in two dimensions to allow to depict all
    curves on one diagram.)
\label{fig_sc}}
\end{figure}

Figure~\ref{fig_sc} shows the specific heat in two and three dimension
for $n=128$ at $\gamma=0.1, 0.4$, and $1.0$ as a function of
temperature.  For $\gamma=0.1$ and $\gamma=1.0$, there is one peak in
the specific heat which is sharp for $\gamma=0.1$ and relatively broad
for $\gamma=1.0$. This is consistent with the scenario described
above, where at $\gamma=0.1$ there should be a first-order transition
in the thermodynamic limit, while at $\gamma=1.0$ we expect a
transition in the $\theta$-point universality class.  Also as
predicted above, for $\gamma=0.4$ there are two well-formed peaks in
the specific heat. In three dimensions the peak at lower temperature
is sharp, while the one at higher temperatures is relatively broad,
consistent with a $\theta$-like transition from the swollen coil to a
collapsed globule at a moderate temperature, followed by a stronger
globule-crystal transition at a lower temperature. In two dimensions,
there are two peaks of roughly equal height. However, they are not
well separated, which indicates that we need to go to longer lengths
to study these transitions.

\subsection{Low temperature phases}

Before considering the order of the phase transitions, especially the
globule-to-crystal transition, we verify that the low-temperature
phases have the properties assumed above. In particular, we
demonstrate that while the globular phase displays no orientational
order, the phase for large $\beta_{hb}$ at fixed $\beta_{nh}$ is a
crystal phase which displays strong orientational order by showing
that in this phase the bonds between monomers prefer to align with one
axis of the lattice.

\begin{figure}[ht!]
  \centering
  \includegraphics[width=0.7\textwidth]{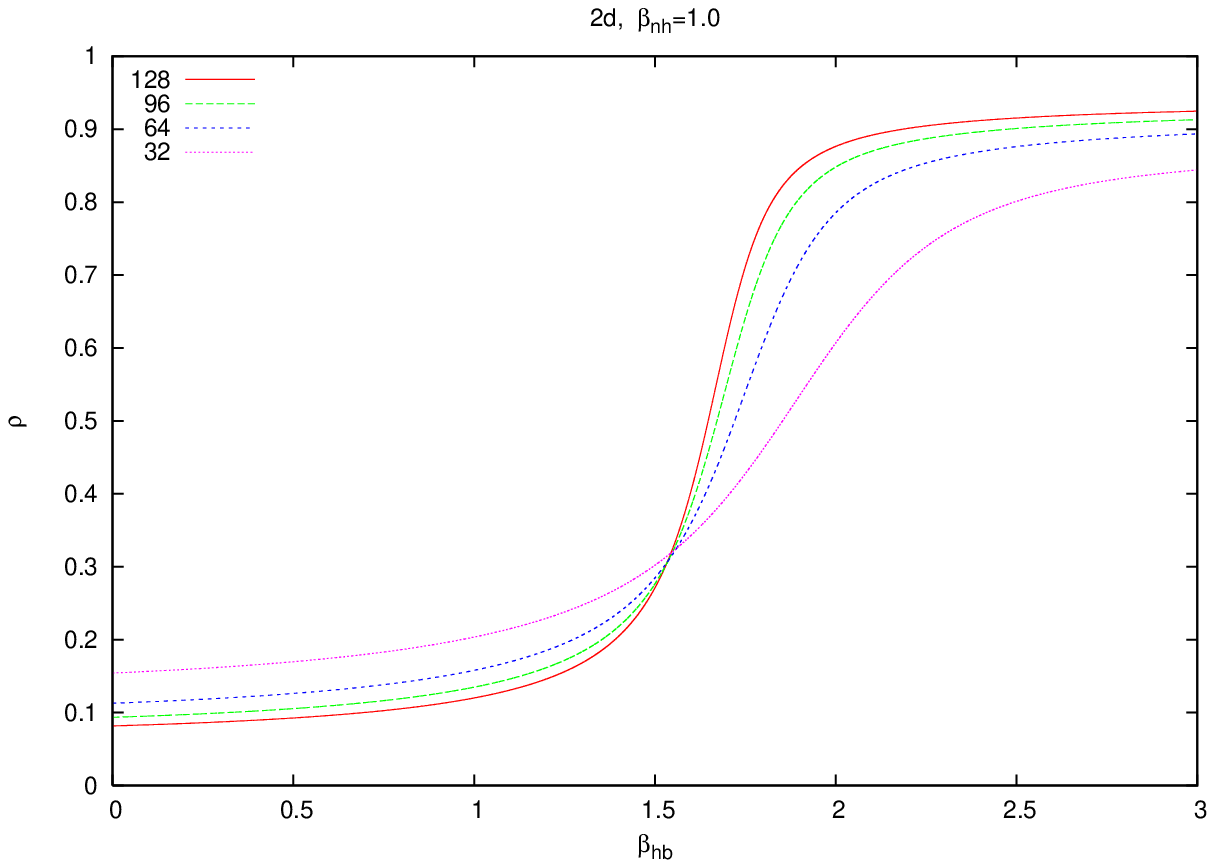}
  \includegraphics[width=0.7\textwidth]{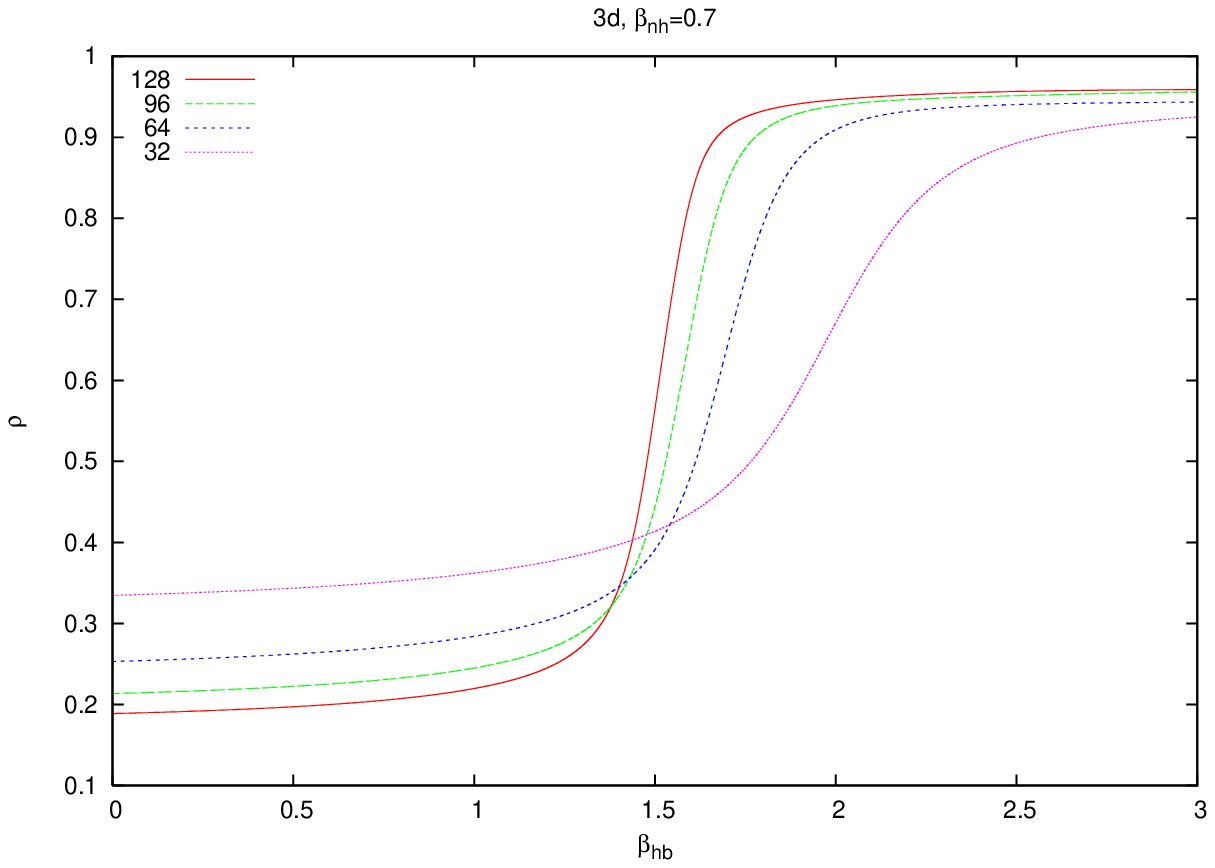}
  \caption{The plots display the anisotropy parameter $\rho$ of the system
  in two and three dimensions.
\label{fig_anisotropy}}
\end{figure}

To detect orientational order, we utilize an anisotropy parameter
defined in Bastolla and Grassberger \cite{bastolla:1997-01}. If we
denote the number of bonds parallel to the $x$-, $y$-, and $z$-axes by
$n_x$, $n_y$, and $n_z$, respectively, we define
\begin{equation}
\rho=1.0-\frac{\min(n_x,n_y,n_z)}{\max(n_x,n_y,n_z)}\;.
\end{equation}
In a system without orientational order, this quantity tends to zero
as the system size increases.  A non-zero limiting value less than one
of this quantity indicates weak orientational order with
$n_{min}\propto n_{max}$, while a limiting value of one indicates
strong orientational order, where $n_{max}\gg n_{min}$.

We consider a fixed value of $\beta_{nh}$ such that the system is
collapsed for any value of $\beta_{hb}$.  For small values of
$\beta_{hb}$ the polymer is in the globular phase, while for large
values it is expected to be in the crystal phase, see
Figure~\ref{fig_lambda}. In two dimensions, we use $\beta_{nh}=1.0$,
while in three dimensions we use $\beta_{nh}=0.7$.

Figure~\ref{fig_anisotropy} shows $\rho$ as a function of $\beta_{hb}$
for different lengths ranging from $32$ to $128$ in two and three
dimensions.  For small $\beta_{hb}$, we find that $\rho$ converges to
zero as $n^{-1/2}$ as expected if only statistical fluctuations are
present. Similarly, for large $\beta_{hb}$, we find that $\rho$
converges to one in a corresponding fashion. This indicates the
presence of strong orientational order in the large
$\beta_{hb}$-phase, which we then deduce to be the ordered crystal.
Intriguingly, for two dimensions only, there exists a value of
$\beta_{hb}$, $1.54$, at which $\rho$ seems to be independent of
system size.  Moreover, by choosing an appropriate exponent
($\phi\approx0.75$), one can show indications of a scaling collapse of
the data near this point. This is our first indication that the
two-dimensional and three-dimensional globule-to-crystal transitions
are different.

\subsection{The phase transitions}

\subsubsection{Swollen coil to folded crystal and swollen coil to globule}

As discussed elsewhere \cite{my:2007-01}, when $\gamma=0$, that is
$\beta_{nh}=0$, there is a single phase transition which is
first-order in both two and three dimensions.  We have verified that
for small $\gamma$, for which we choose $\gamma=0.1$, this scenario
remains intact. A bimodal distribution can be clearly seen forming as
the system size becomes larger very strongly in three dimensions, and
more weakly in two dimensions.

To show that the $\theta$-transition extends from $\gamma=1/2$ to
larger values of $\gamma$, we can focus on the case of $\gamma=1$,
which means that $\beta_{hb}=0.0$ and \textit{hb}-interactions are
irrelevant.

\begin{figure}[ht!]
  \centering
  \includegraphics[width=0.7\textwidth]{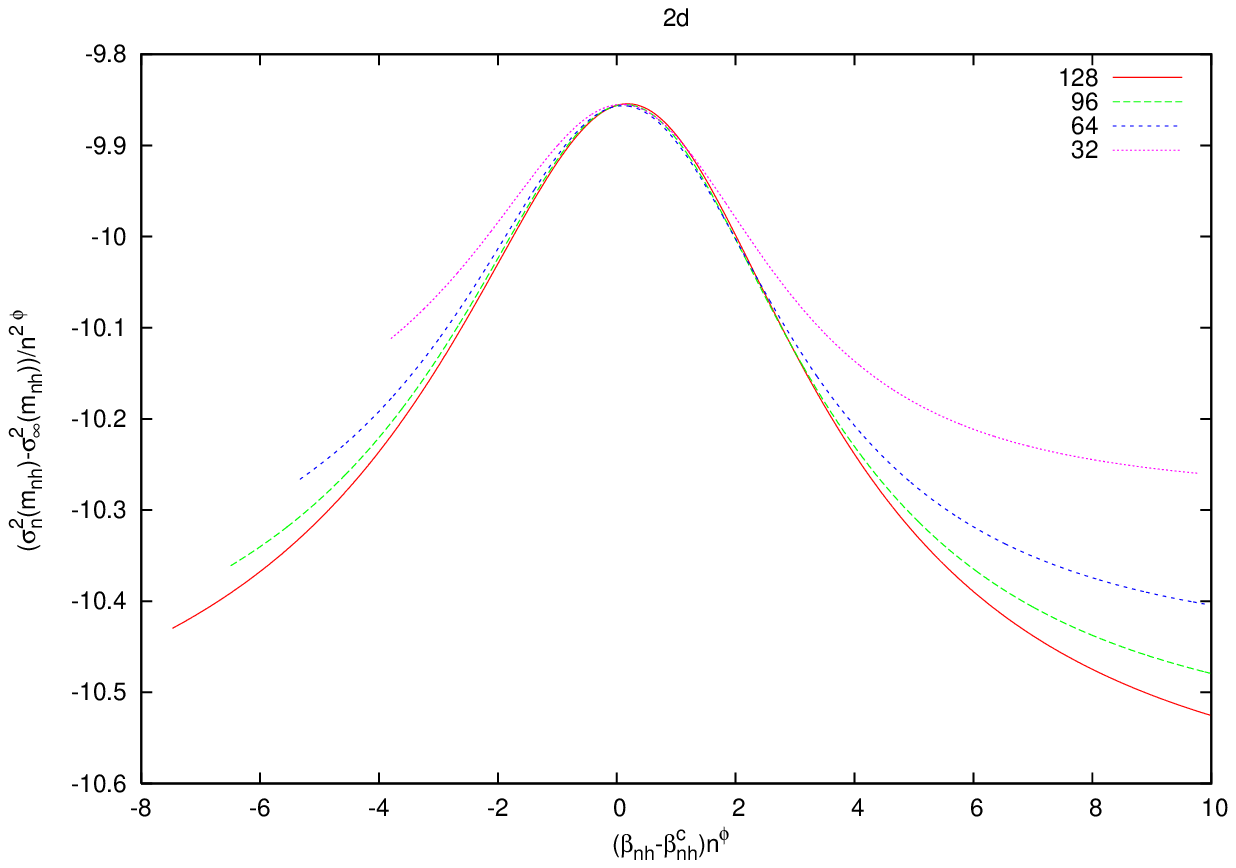}
  \includegraphics[width=0.7\textwidth]{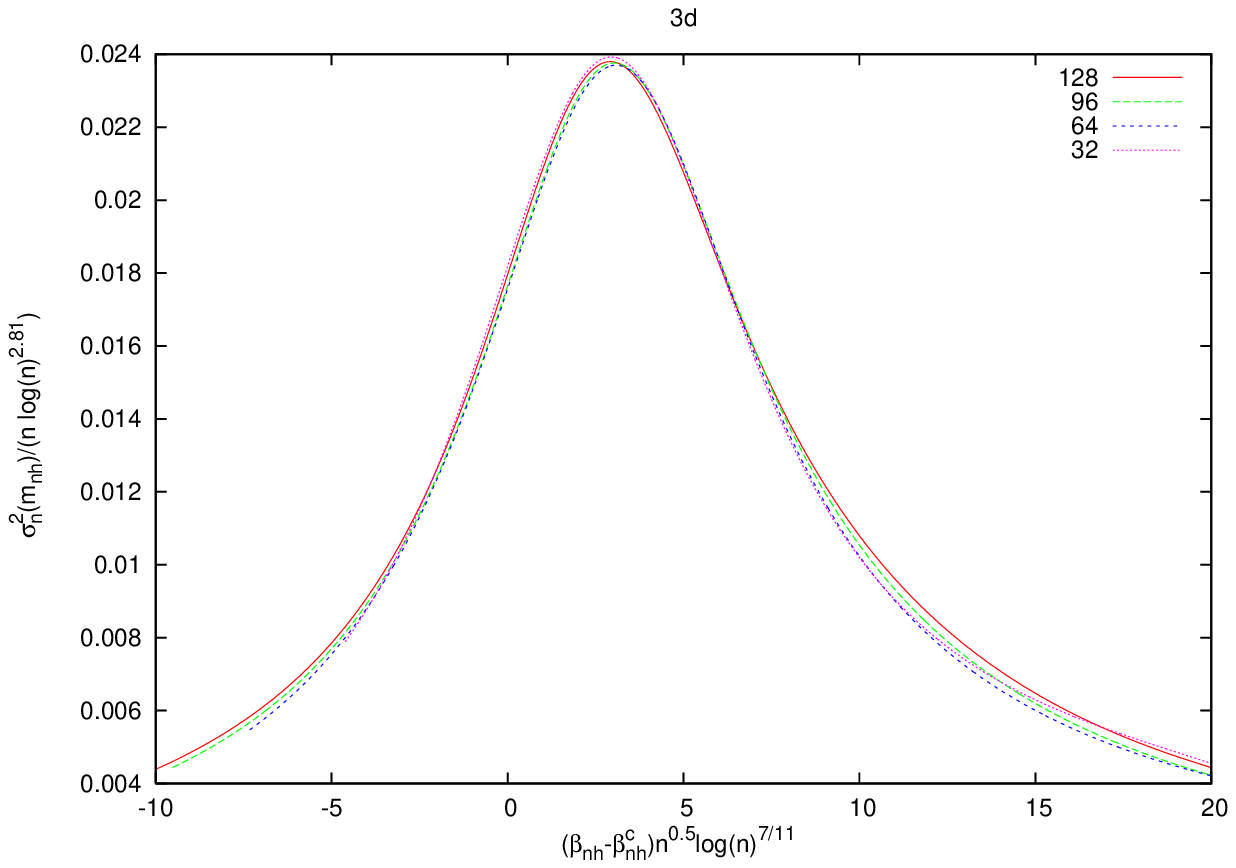}
  \caption{Scaling plots for the fluctuations in $m_{nh}$ for the $\theta$-like
   transitions occuring along the line $\beta_{hb}=0.0$.
\label{fig_scalingbs0.0}}
\end{figure}

In two dimensions the maximum of the fluctuations per monomer for
length $n$ for the $\theta$-point behaves as
\begin{equation}
\sigma_n^2(m_{nh})\sim\sigma_{\infty}^2(m_{nh})+a\cdot n^{2\phi}\;,
\end{equation}
where $\phi=3/7$.  We have estimated from our data collected at short
length that $\phi=0.49$.  This is consistent with the observation that
for finite system size the effective crossover exponent decreases from
a value well above $0.5$ \cite{prellberg:1994-01} to the theoretical
value predicted for the $\theta$-point $\phi=3/7$. Having extrapolated
the limiting value $\sigma_{\infty}^2(m_{nh})$, we show in
Figure~\ref{fig_scalingbs0.0} a scaling plot of the dependence of the
\emph{singular} part of the fluctuations as a function of
$\beta_{nh}$.

In three dimensions \cite{deGennes:book,duplantier:1986} theory
predicts $\phi=1/2$ and a logarithmic divergence of the maximum of
fluctuations. The scaling of the fluctuations around the transition
for $\beta_{hb}=0$ are shown in Figure~\ref{fig_scalingbs0.0}.  At
short lengths, we find strong corrections to scaling, in accord with
the observations in \cite{grassberger:1995-01,hager:1999}.  The
effective exponent of the logarithm is equal to $2.8$, which is about
an order of magnitude more than the value $3/11$ predicted
\cite{duplantier:1986}.  However we have checked that the exponent
decreases with the system size.

\subsubsection{Collapsed globule to folded crystal}

\begin{figure}[ht!]
  \centering
  \includegraphics[width=0.7\textwidth]{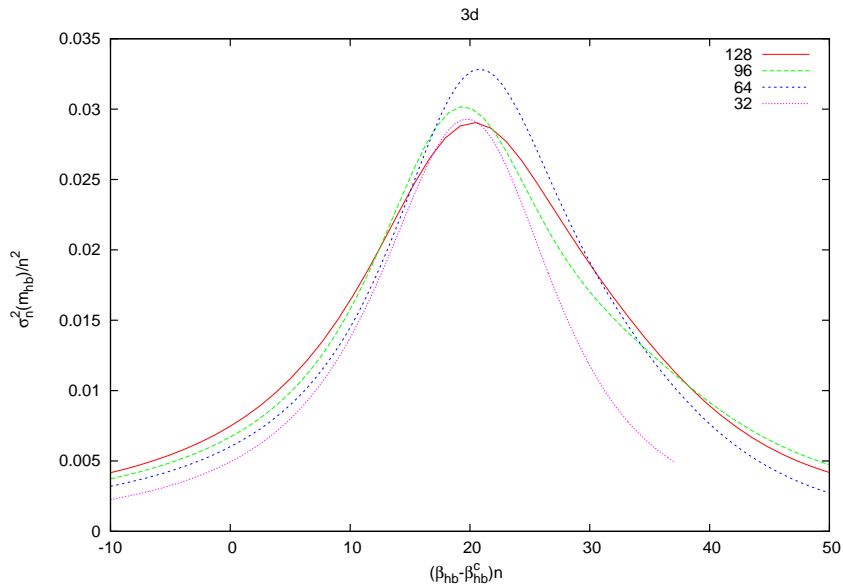}
  \caption{A scaling plot for the fluctuations in $m_{hb}$ for the
   globule-to-crystal transition occuring along the line
   $\beta_{nh}=0.7$ in three dimensions.
 \label{fig_scaling}}
\end{figure}

We begin with the three-dimensional case and return to our simulations
at fixed $\beta_{nh}=0.7$.  On varying $\beta_{hb}$ we find a
first-order transition from the globule to the crystal. The maximum of
fluctuations per monomer $\sigma^2(m_{hb})/n$ increases linearly in
$n$, and the shift of the inverse temperature scales as $1/n$, in
accord with finite-size scaling of a first-order
transition. Figure~\ref{fig_scaling} shows the corresponding scaling
collapse, with an extrapolated value of $\beta_{hb}^c=1.34$.

We turn to our simulations of the two-dimensional case at fixed
$\beta_{nh}=1.0$.  Now, on varying $\beta_{hb}$ we find a transition
which is much stronger than the $\theta$-point, but shows no
indication of being first-order: the maximum of fluctuations per
monomer $\sigma^2(m_{hb})/n$ diverges with an exponent less than one.
A scaling plot using a consistent power law is not convincing. Our
best estimate for the crossover exponent comes in fact from the
scaling of the anisotropy parameter $\rho$ discussed above, which
gives $\beta_{hb}^c=1.54$ and a crossover exponent in the vicinity of
$\phi=0.75$. The scaling of the fluctuations is not inconsistent with
these values.

\subsection{Summary}

\begin{figure}[ht!]
  \centering
  \includegraphics[width=0.7\textwidth]{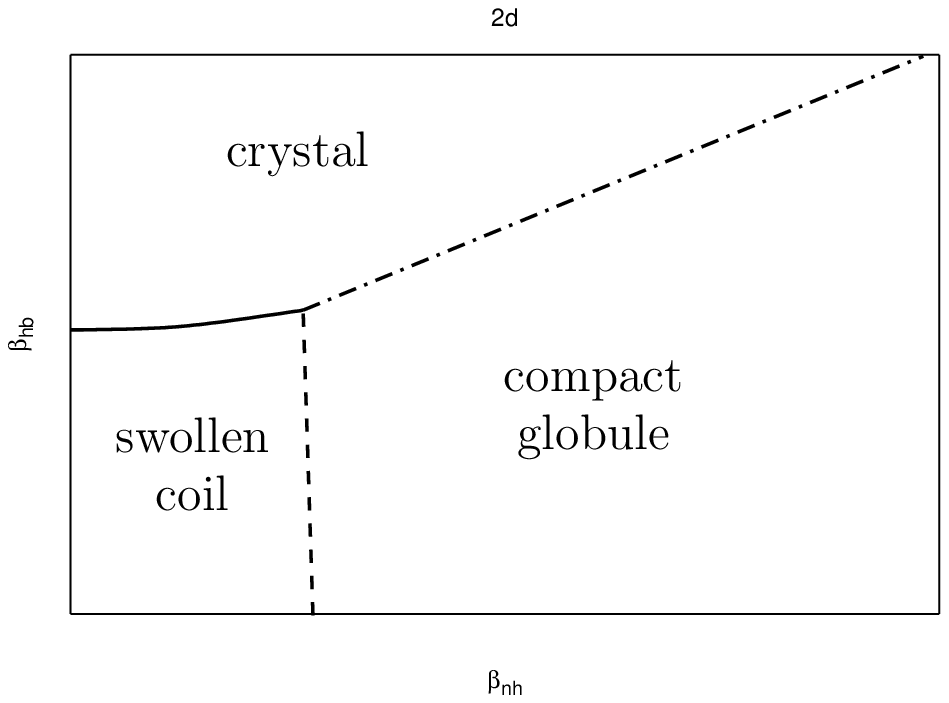}
  \includegraphics[width=0.7\textwidth]{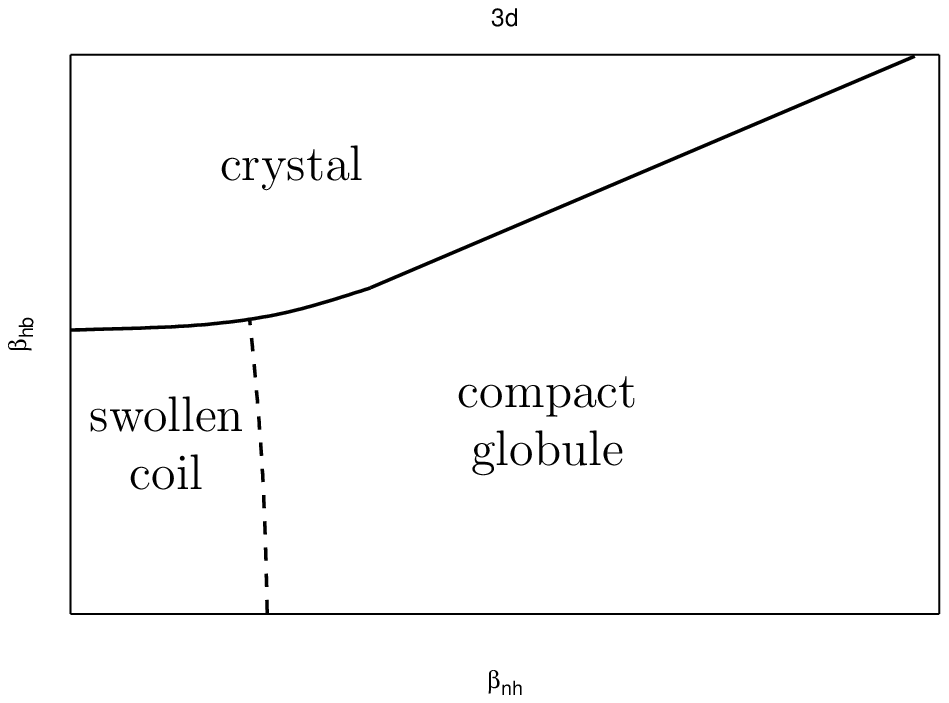}
  \caption{The schematic conjectured phase diagrams in two and three
    dimensions. A solid line denotes a first-order transition, and a
    dashed line a $\theta$-like transition. The dot-dashed line
    represents a putative second-order phase transition that is not
    yet completely characterised.}
  \label{fig_pdschemata}
\end{figure}

We summarize our findings by presenting conjectured phase diagrams in
two and three dimensions in Figure~\ref{fig_pdschemata}.  For large
values of the ratio of the interaction strength of hydrogen-bonds to
non-hydrogen bonds, a polymer will undergo a single first-order phase
transition from a swollen coil at high temperatures to a folded
crystalline state at low temperatures.  On the other hand, for any
ratio of these interaction energies less than or equal to one there is
a single $\theta$-like transition from a swollen coil to a liquid
droplet-like globular phase. For intermediate ratios two transitions
can occur, so that the polymer first undergoes a $\theta$-like
transition on lowering the temperature, followed by a second
transition to the crystalline state. In three dimensions we find that
this transition is first order, while in two dimensions we find that
this transition is probably second order with a divergent specific
heat. It can be argued using a zero-temperature argument that this
scenario exists as soon as the ratio of the interaction energies is
greater than one. In this way the phase diagram described is
qualitatively similar to the one of the semi-stiff interacting polymer
model described in three dimensions by Bastolla and Grassberger
\cite{bastolla:1997-01}.

The interesting questions that remain for future work include further
characterizing the globule-to-crystal transition in two dimensions and
also clearly delineating the range of interaction ratios for which two
transitions occur.

\vspace{1em}
 
Financial support from the Australian Research Council and the Centre
of Excellence for Mathematics and Statistics of Complex Systems is
gratefully acknowledged by the authors.  All simulations were
performed on the computational resources of the Victorian Partnership
for Advanced Computing (VPAC).

\end{document}